\documentclass{article} 

\usepackage[latin1]{inputenc}
\usepackage{a4wide}
\usepackage{amsmath}
\usepackage{graphics}

\makeatletter

\begin{document}

\bigskip

\title{\Huge \bf A high-velocity black hole on a Galactic-halo orbit
  in the solar neighborhood}

\author{I.F. Mirabel$^{1,2,*}$, V. Dhawan$^{3,*}$, R.P. Mignani$^{4}$,
I. Rodrigues$^{1}$, F. Guglielmetti$^{5}$} 

\maketitle

\bigskip
\centerline{To be published in Nature, vol. 413 pages 139-141 (13 September 2001)} 
\bigskip

\noindent$^1$Service d'Astrophysique / CEA, CE-Saclay. 91191 
Gif/Yvette, France. (fmirabel@cea.fr)

\noindent$^2$Instituto de Astronom\'\i a y F\'\i sica del
Espacio. cc5, 1428 Bs As, Argentina

\noindent$^3$National Radio Astronomy Observatory. P.O. Box 0,
Socorro, New Mexico 87801, USA.  (vdhawan@aoc.nrao.edu)

\noindent$^4$European Southern Observatory. Karl-Schwarzschild-Strasse
2, Garching bei Munchen, DE-85740, Germany. (rmignani@eso.org)

\noindent$^5$Space Telescope Science Institute.  3700 San Martin
Drive. Baltimore.  MD 21218 USA. (fabrizia@stsci.edu)

\noindent$^*$These authors contributed equally to this work

\bigskip

\noindent {\LARGE \bf \dotfill }

\bigskip

\noindent{\bf Only a few of the dozen or so stellar-mass black holes 
  have been observed away from the plane of the Galaxy$^1$.  Those few
  could have been ejected from the plane as a result of a ``kick''
  received during a supernova explosion, or they could be remnants of
  the population of massive stars formed in the early stages of
  evolution of the Galaxy.  Determining their orbital motion should
  help to distinguish between these options. Here we report the
  transverse motion (in the plane of the sky) for the black hole X-ray
  nova XTE J1118+480 (refs 2-5), from which we derive a large space
  velocity. This X-ray binary has an eccentric orbit around the
  Galactic Centre, like most objects in the halo of the Galaxy, such
  as ancient stars and globular clusters. The properties of the system
  suggest that its age is comparable to or greater than the age of the
  Galactic disk. Only an extraordinary ``kick'' from a supernova could
  have launched the black hole into an orbit like this from a birth
  place in the disk of the Galaxy.}

\bigskip

The high galactic latitude ({\it l} = 157.7$^{\circ}$, {\it b} =
+62.3$^{\circ}$) X-ray nova XTE J1118+480 was discovered$^2$ with the
RXTE All-Sky Monitor on 2000 March 29.  It exhibited slow outbursts
that lasted $\sim$7 months with a peak X-ray luminosity of 4 x
10$^{35}$ (D/kpc)$^2$ erg s$^{-1}$ and energy spectra typical of black
hole binaries in low/hard state$^6$. From observations of the $\sim$19
mag optical counterpart$^3$ in quiescence a mass function for the
compact object f(M) $\sim$ 6.0$\pm$0.4 M$_{\odot}$ and an average distance of
$\sim$ 1.85$\pm$0.36 kpc were determined$^{4,5}$. For $\sim$100 days
the source exhibited a steady and slowly variable unresolved radio
counterpart$^{7,8}$ with persistent inverted radio spectrum, which is
interpreted as optically thick emission from a compact, powerful
synchrotron jet$^{8,9}$, that could have a size $\leq$0.03 AU$^9$.

\vskip .1in

To measure the transverse
motion on the plane of the sky of XTE J1118+480 we carried out
observations of the radio counterpart 
with the Very Long Baseline Array (VLBA) at 15.4 GHz
($\lambda$2 cm) and 8.4 GHz ($\lambda$3.6 cm) on 2000 May 4 and July
24; afterwards it faded below detection.
In both epochs the source was unresolved by the synthesized beams of
1$\times$0.6 milli arc sec (mas) and 2$\times$1 mas respectively, 
which correspond to physical
dimensions smaller than $\sim$0.7 D/kpc AU, and a brightness
temperature $\geq$10$^9$ K. Because of the high galactic latitude and
low column of interstellar gas along the line of sight (N$_H$ $\sim$
10$^{20}$ cm$^{-2}$)$^{10}$, the VLBA images are relatively unaffected by Galactic electron scattering, allowing us to determine in each
epoch the position of the compact, unresolved radio source with
relative errors $\leq$0.35 mas.

The observations were done in cycles that included the target, the
strong calibrator 3C273, a primary calibrator (Cal. 1) used as
phase-reference source, and a second extragalactic calibrator (Cal.2).  
Their high elevation and good weather ensured that the phase
connection was successful.  The epoch, source, frequency, bandwidth,
position, and flux for the VLBA observations of XTE J1118+480 and the
two extragalactic sources used as references for relative astrometry
are given in Table 1. The target had inverted and flat spectra, with
similar fluxes within 10\% of those interpolated from the monitoring
with the Ryle$^7$ and the Very Large Array telescopes$^{11}$, which
confirms that the source was unresolved by the VLBA synthesized beams. 
Figure 1 shows a drift relative to the extragalactic frame between the
two epochs of observation that after correction for parallax amounts
to a proper motion on the plane of the sky of 18.3$\pm$1.6 mas 
yr$^{-1}$ with position angle = 246$^{\circ}$$\pm$6$^{\circ}$. 

To   obtain a  further assessment of the proper motion, we  have
performed an  independent measurement  in the  optical using
the photographic plates of the Palomar Observatory Sky Survey 
(POSS I \& II), digitized for the Guide Star Catalogue-II project, 
which cover a time span of about 43 years.
By relative astrometry we find that the secondary has a proper
motion of 12.0$\pm$3.2 mas yr$^{-1}$ with a PA = 
240$^{\circ}$$\pm$15$^{\circ}$.   Such a value is consistent
both in magnitude and position angle with the more precise measurement
obtained at radio wavelengths (see Figure 1).

\vskip .2in

The velocity components U, V, and W directed to the Galactic centre,
rotation direction, and north Galactic pole are derived using the
equations of transformation$^{12}$, and assuming the sun moves
(U$_{\odot}$,V$_{\odot}$,W$_{\odot}$) = (9, 12, 7) km s$^{-1}$
relative to the local standard of rest (lsr)$^{13}$.  For the reasons
stated above, in the following we use the proper motion measured with 
the VLBA which
after correction for change in parallax is: $\Delta$$\alpha$ = -
16.8$\pm$1.6 mas yr$^{-1}$ and $\Delta$$\delta$ = -
7.4$\pm$1.6 mas yr$^{-1}$. Distances of 1.8$\pm$0.6 kpc$^4$ and
1.9$\pm$0.4 kpc$^5$ are reported for XTE J1118+480, but for the radial
velocity of the centre of mass somewhat discrepant values are
estimated$^{4,5}$: + 26$\pm$17 km s$^{-1}$ and - 15$\pm$10 km
s$^{-1}$, respectively.  For a mean value d = 1.85$\pm$0.36 kpc and V$_r$ = - 15
$\pm$ 10 km s$^{-1}$, we find (U = - 97 $\pm$23, V = - 101 $\pm$23, W
= - 39 $\pm$12) km sec$^{-1}$ in the lsr frame. This implies that the
source moves away from the Galactic centre, has slower rotation about
the Galactic centre than the stars in the disk, and a 3$\sigma$ motion
towards the Galactic plane. Adopting V$_r$ = + 26$\pm$17 km s$^{-1}$,
we obtain essentially similar results (U = - 114 $\pm$24, V = - 94
$\pm$23, W = - 2 $\pm$17) km sec$^{-1}$, but with no significant
motion perpendicular to the Galactic plane.  Within the range of
uncertainties for the distance (1.4-2.4 kpc) and radial velocity (-15
to +26 km s$^{-1}$), and irrespective of the specific values adopted,
according to the definition of ``high velocity star in the solar
neighborhood'' --[(W+10) $\geq$ 30 km s$^{-1}$ and/or (U$^2$ +
V$^2$)$^{1/2}$ $\geq$ 65 km s$^{-1}$]$^{14}$-- XTE J1118+480 can be considered  
a high velocity X-ray binary, since the velocity relative to the lsr 
is 145 km s$^{-1}$. The V component of XTE
J1118+480 implies low rotation (V$_{\phi}$ $\sim$ 122 km s$^{-1}$)
about the Galactic centre, contrary to the Galactic disk population,
which in the solar neighborhood is in rapid rotation of $\sim$220 km
s$^{-1}$.

\vskip .2in

Figure 2 shows the orbit of the black hole binary about the
Galactic centre derived from the mean values (U = - 105$\pm$16, V = -
98$\pm$16, W = - 21$\pm$10) km sec$^{-1}$ in the lsr frame, a distance
of 1.85$\pm$0.36 kpc from the Sun, and the standard galactic
gravitational potential that includes the disk, bulge and halo 
components$^{15}$. A change of 10\% 
in any of the free parameters of the standard galactic gravitational 
potential$^{15}$ results in changes of less than 5\% in any of the 
orbital parameters of XTE J1118+480.

\vskip .3in

We now discuss the possibility that XTE J1118+480 could have been
launched from the Galactic plane into its halo orbit by a supernova
explosion that took place during the formation of the black hole.
Using the properties of XTE J1118+480$^{4,5}$ 
(M$_{BH}$=6.9$\pm$0.9 M$_{\odot}$, mass of the binary companion donnor star 
M$_{donor}$=0.3$\pm$0.2 M$_{\odot}$, binary separation of 3
R$_{\odot}$ in circular orbit with orbital period of 0.17 days), and the equations that
describe the impulse from symmetric explosions$^{16}$, it is found that 
to accelerate the black hole up to a peculiar velocity
of 217 km s$^{-1}$  by a symmetric
explosion more than 40 M$_{\odot}$ would be suddenly
ejected during the stellar collapse, which is implausibly large. 

Alternatively, the supernova explosion could be asymmetric and 
the collapsar receive an additional kick that can be assumed to be
equal for both black holes and neutron stars, the runaway velocity
being proportional to the inverse of the mass$^{17}$. Comparing with
the runaway velocities of neutron stars, the momentum
of XTE J1118+480 is similar to that of a solitary neutron star
moving at 1000 km s$^{-1}$, $\sim$10 times the average and $\geq$3
times the largest linear momentum among millisecond pulsars$^{18}$.
Therefore, an origin in the galactic disk would imply that 
XTE J1118+480 received the most extreme natal 
impulse among the known binaries that contain compact objects. 
Because binaries with more massive components 
are more likely to remain bound after the kicks$^{17}$, and the 
binary orbit can circularize by tidal friction in few Myrs$^{19}$, a disk 
origin through an extraordinary explosion cannot be ruled out.  

\vskip .2in

Instead of being formed in the Galactic disk the 
black hole in XTE J1118+480 may be the relic of an ancient massive 
star formed in the Galactic halo. The values of U and V are consistent  
with the large random motions of old halo stars flying through 
the solar neighborhood$^{20}$ that have metallicities
[Fe/H] = -1.2 to -1.4. Furthermore, the Galactic orbit shown in Figure 2 is similar to that of some 
globular clusters (e.g. NGC 6656 with [Fe/H] = -1.7)$^{21}$.  
We point out that there are two additional observations that are consistent 
with the hypothesis of a halo origin: 1) the very 
low metallicity Z/Z$_{\odot}$$\sim$0.1 of the 
reflector medium derived from broad band X-ray spectroscopy$^{22}$, 
and 2) the depletion of carbon and enhancement of nitrogen found in the  
ultraviolet spectrum with HST, which would require a long lasting 
Carbon-Nitrogen-Oxigen process, suggesting that the secondary 
lost a large fraction of 
its mass and is presently exposing the layers that were 
originally below the surface$^{23}$. Therefore, XTE J1118+480 
may be contemporary of globular clusters, being one of the 
10$^4$-10$^5$ black holes that were ejected from these 
old stellar systems$^{24}$ and at present swirl around in the halo. 

\vskip .2in

In the year 2000 XTE J1118+480 brightened
from quiescence by $\sim$6 mag. Our analysis of the historical
optical plates reveals that  in the years 1995 and 1996 it had 
optical outbursts of $\sim$2 mag that passed without notice, although could 
have been easily measured. We point out that XTE J1118+480 was a black 
hole nova with a remarkably large
optical-to-X-ray flux ratio$^{10}$ that reached a peak luminosity of 
about 10$^{36}$ erg s$^{-1}$ in the 1-160 keV band$^6$. Presently it is
close to the Sun, but at the distance of the Galactic centre it
would have not been detected in most
surveys with X-ray instruments of large field of view. Therefore, it
is possible that many faint X-ray binaries have been
missed, which like XTE J1118+480 may sporadically appear as
microquasars$^{25}$, namely as sources of collimated beams of relativistic
particles and high energy photons. The association of these
microquasars with a subset of unidentified gamma-ray
sources$^{26}$ that are rather variable, soft, faint, and have
-as globular clusters- a scale-height above the galactic plane of
$\sim$2 kpc, is an intriguing possibility.

\vfill\eject

\centerline{\bf References}

\vskip .1in

\noindent$^1$White, N. E. \& van Paradijs J., The Galactic
Distribution of Black Hole Candidates in Low-Mass X-ray Binary Systems.
Astrophys. J. {\bf 473}, L25-L29 (1996). 

\noindent$^2$Remillard, R., Morgan, E., Smith, D. \& Smith, E. XTE
J1118+480.  IAU Circ. No 7389 (2000).

\noindent$^3$Uemura, M., et al. Discovery and Photometric Observation
of the Optical Counterpart in a Possible Halo X-ray Transient, XTE
J1118+480.  Pub. Astron. Soc. Japan, {\bf 52}, L15-L19 (2000).

\noindent$^4$McClintock, J.E., Garcia, M.R., Caldwel, N., Falco, E.E.,
Garnavich, P.M. \& Zhao, P. A Black Hole of greater than 6 M$_{\odot}$ in
the X-ray Nova XTE J1118+480. Astrophys. J. {\bf 551}, L147-150 (2001).

\noindent$^5$Wagner, R.M., Foltz, C.B., Shahbaz, T., Casares, J.,
Charles, P.A., Starrfield, S.G., \& Hewett, P.  The Halo Black Hole
X-ray Transient XTE J1118+480.  Astrophys. J. {\bf 556}, 42-46 (2001).

\noindent$^6$McClintock, J.E. et al. Complete Simultaneous Spectral 
Observations of
the Black-Hole X-ray Nova XTE J1118+480. Astrophys. J. {\bf 555}, 477-482 (2001).

\noindent$^7$Pooley, G.G., Waldram, E.M. XTE J1118+480. IAU Circ. No
7390 (2000).

\noindent$^8$Dhawan, V., Pooley, G.G., Ogley, R.N., Mirabel, I.F. XTE
J1118+480. IAU Circ. No 7395 (2000).

\noindent$^9$Fender, R.P., Hjellming, R.M., Tilanus, R.P.J., Pooley,
G.G., Deane, J.R., Ogley, R.N. \& Spencer, R.E. Spectral Evidence for
a Powerful Compact Jet from XTE J1118+480.  Mon. Not. R. Astron. Soc.
{\bf 322}, L23-27 (2001).

\noindent$^{10}$Hynes, R.I., Mauche, C. W., Haswell, C. A., Shrader,
C. R.; Cui, W., Chaty, S. The X-Ray Transient XTE J1118+480:
Multiwavelength Observations of a Low-State Minioutburst. Astrophys.
J. {\bf 539}, L37-L40 (2000).

\noindent$^{11}$Dhawan, V., Pooley, G.G., Ruppen, M., Hjellming, R.H.,
Ogley, R.N., Mirabel, I.F. Radio Monitoring of XTE J1118+480. In
preparation (2001).

\noindent$^{12}$Johnson, D.R.H., Soderblom, D.R. Calculating Galactic
Space Velocities and their Uncertainties. Astron. J. {\bf 93}, 864-867
( 1987)

\noindent$^{13}$Mihalas, D., Binney, J. {\it Galactic Astronomy} (San
Francisco: Freeman). pages 382-384 (1981).

\noindent$^{14}$Oort, J.H. Asymmetry in the distribution of stellar
velocities. The Observatory {\bf 49}, 302-304 (1926).

\noindent$^{15}$Johnston, K.V., Hernquist, L., \& Bolte, M.  Fossil
Signatures of Ancient Accretion Events in the Halo. Astrophys. J. {\bf
  465}, 278-287 (1996).

\noindent$^{16}$Nelemans, G., Tauris, T.M., van den Heuvel, E.P.J. 
Constraints on mass ejection in black hole formation derived from
black hole X-ray binaries. Astron. Astrophys. {\bf 352}, L87-L90
(1999).

\noindent$^{17}$Fryer, C.L., \& Kalogera, V. Theoretical black hole
mass distributions. Astrophys. J. {\bf 554}, 548-560 (2001).

\noindent$^{18}$Toscano, M., Sandhu, J. S., Bailes, M., Manchester,
R. N., Britton, M. C., Kulkarni, S. R., Anderson, S. B., Stappers, B.
W. Millisecond pulsar velocities. Mon. Not. R. Astron. Soc. {\bf 307},
925-933 (1999).

\noindent$^{19}$Zahn, J.-P. Tidal friction in close binary stars. 
Astron. Astrophys. {\bf 57}, 383-394 (1977).

\noindent$^{20}$Chiba, M., Beers, T.C. Kinematics of Metal-Poor Stars
in the Galaxy.  III. Astron. J.  {\bf 119}, 2843-2865 (2000).

\noindent$^{21}$Dauphole, B., Geffert, M., Colin, J., Ducourant, C.,
Odenkirchen, M., Tucholke, H.-J. The kinematics of globular clusters,
apocentric distances and a halo metallicity gradient.  Astron.
Astrophys. {\bf 313}, 119-128 (1996).

\noindent$^{22}$ Frontera, F. et al.  A measurement of the broad-band
spectrum of XTE J1118+480 with BeppoSAX and its astrophysical
implications. Astrophys. J. in press. (2001) .

\noindent$^{23}$Haswell, C.A. et al. XTE J1118+480. 
IAU Circ. No 7407 (2000).

\noindent$^{24}$Kulkarni, S.R., Hut, P., McMillan, S. Stellar black
holes in globular clusters. Nature {\bf 364}, 421-423 (1993).

\noindent$^{25}$Mirabel, I.F. \& Rodr\'\i guez, L.F. Microquasars in our
Galaxy. Nature, {\bf 392}, 673-676 (1998).

\noindent$^{26}$Grenier, I.A. EGRET Unidentified Sources. {\it GeV-TeV
  Gamma Ray Astrophysics Workshop : towards a major atmospheric
  Cherenkov detector VI}. Eds. B.L. Dingus, M.H. Salamon, and D. B.
Kieda. (Melville, N.Y.): American Institute of Physics {\bf 515},
261-279 (1999).

\vskip .2in 

{\bf Acknowledgements}

We thank G. Pooley for
information on the fluxes observed with the Ryle telescope, A. Spagna 
for providing his astrometric software for the optical measurements, 
and R. Fern\'andez 
for help with software programming.  I.F.M.  
thanks J. Paul, J.-P. Zahn,  R.M. Wagner, G. Israelian, R. Rebolo, E. Ergma 
and A. King for helpful comments and discussions.  
The National Radio Astronomy Observatory is a facility of the National
Science Foundation operated under cooperative agreement 
by Associated Universities, Inc.  The Guide Star Catalogue-II is a 
joint project  of the Space Telescope Science Institute and the 
Osservatorio  Astronomico di Torino with additional support provided  
by the European Southern Observatory, Space Telescope European Coordinating Facility, the International  GEMINI project and  the 
European Space Agency Astrophysics Division. I.F.M. is a member of the 
Consejo Nacional de Investigaciones Cient\'\i ficas y T\'ecnicas of  
Argentina, and I.R. Fellow of the Conselho Nacional de Desenvolvimento 
Cient\'\i fico e Tecnol\'ogico of Brazil.

\clearpage

\begin{table}
\begin{center}
TABLE 1. VLBA astrometry of XTE J1118+480.\\
\vspace{0.5cm}
\begin{tabular}{c|l|c|c|c|c|r} \hline \hline
\label{VLBA}
 & Source & Freq  & BW    & RA(J2000) & DEC(J2000) &  Flux~ \\
      &        & (GHz) & (MHz) & (h  m  s $\pm$ mas)  & (d  '  " $\pm$
      mas) &  (mJy)~ \\ \hline
   &               &           &                &               &  \\
 &            &           &                &               &  \\

 A & X1118 & 8.4  & 16 & 11 18 10.7918249$\pm$0.093 & 48 02 12.316177$\pm$0.131 & 5.9$\pm$0.3     \\ 

 A & Ref.2 & 8.4  & 16 & 11 26 57.6550573$\pm$0.014 & 45 16 06.284006$\pm$0.023 & 126$\pm$3.8\\

 A & Ref.1 & 8.4  & 16 & 11 10 46.3458105$\pm$0.232 & 44 03 25.925163$\pm$0.240 & 264$\pm$5.3\\ \hline

 A & X1118 & 15.4 & 32 & 11 18 10.7918131$\pm$0.084 & 48 02 12.316161$\pm$0.111 & 8.1$\pm$0.3 \\  

 A & Ref.2 & 15.4 & 32 & 11 26 57.6550672$\pm$0.032 & 45 16 06.283518$\pm$0.050 & 68$\pm$2.0\\ 

 A & Ref.1 & 15.4 & 32 & 11 10 46.3458105$\pm$0.232 & 44 03 25.925163$\pm$0.240 & 241$\pm$4.8\\ \hline

   &       &           &                &               &  \\
   &           &    &                &               &  \\

 B & X1118 & 8.4  & 64 & 11 18 10.7914440$\pm$0.115 & 48 02 12.313919$\pm$0.155 & 2.7$\pm$0.1 \\

 B & Ref.2 & 8.4  & 64 & 11 26 57.6550224$\pm$0.008 & 45 16 06.283783$\pm$0.014 & 120$\pm$3.2\\

 B & Ref.1 & 8.4  & 64 & 11 10 46.3458105$\pm$0.232 & 44 03 25.925163$\pm$0.240 & 267$\pm$5.2\\ \hline

 B & X1118 & 15.4 & 64 & 11 18 10.7914531$\pm$0.159 & 48 02 12.314120$\pm$0.191 & 2.9$\pm$0.1 \\

 B & Ref.2 & 15.4 & 64 & 11 26 57.6550612$\pm$0.007 & 45 16 06.283338$\pm$0.014 & 65$\pm$1.8\\ 

 B & Ref.1 & 15.4 & 64 & 11 10 46.3458105$\pm$0.232 & 44 03 25.925163$\pm$0.240 & 229$\pm$4.6\\ \hline

\end{tabular}
\end{center}

\end{table}

{

%  \em

\bf{NOTES:} \rm

 A = MJD 51668, 2000 May 04 UT 03:30-07:40; B = MJD 51749, 2000 July
 24 UT 20:00-01:50. Between these two epochs the observed change in
 position of XTE J1118+480 with respect to the extragalactic frame, was
 $\Delta$RA~=~-3.715~mas and $\Delta$Dec~=~-2.149~mas. The positions
 at 8.4 and 15.4~GHz were averaged at each epoch and then differenced. 
 The position of Ref.1 with associated absolute error is from
 the new catalog of VLBA calibrators (Personal Communication from
 D. Gordon, C. Ma, L. Petrov, A. Beasley, E. Fomalont \& A. Peck, in
 preparation). The errors for other sources are formal fitting
 errors. The position of Ref.2 in this table agrees within
 2$\sigma$ with its absolute position from the new catalog. The
 systematic errors tend to cancel when the two epochs are differenced,
 and the residual errors in the relative astrometry are smaller than
 the short-term atmospheric ``seeing" (fluctuations in position caused
 by residual tropospheric phase errors in the phase-referencing
 step). The correlator model used for both epochs was the same and
 contains a seasonally averaged global troposphere, 
 but no short-term weather information. The phase referencing reduces 
 the tropospheric errors to the measured level of 0.3 mas in each axis, 
 as found by breaking the
        data into 1hr intervals and measuring the rms position scatter
        on the source ref.2.  The errors for XTE J1118+480 are 0.35 mas in
        each axis, estimated by adding the (maximum) formal error of
        0.2 mas in quadrature to the 0.3 mas tropospheric errors. 

}

\begin{figure*}[p]
  {\par\centering \resizebox*{1\textwidth}{!}{
      \rotatebox{-90}{\includegraphics{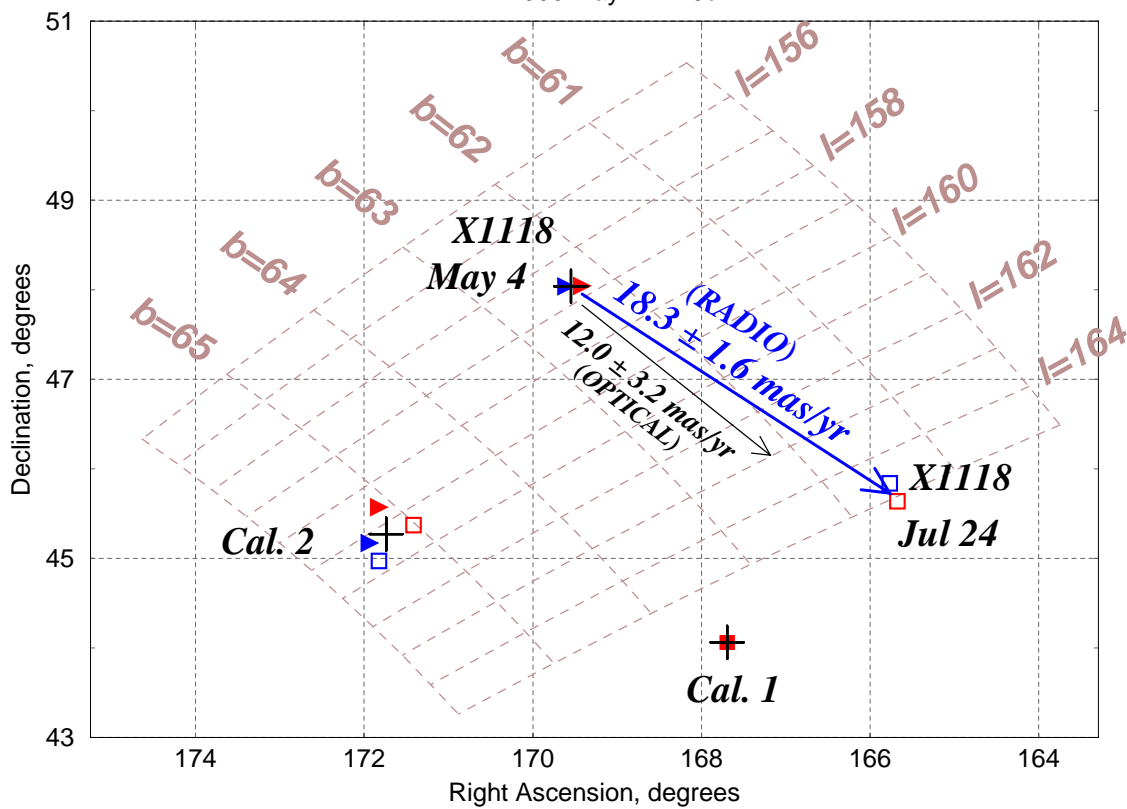}}}
    \par}
  \caption{Proper motion of XTE J1118+480, observed with the VLBA on 
    4 May-24 July 2000. Shift in position (blue arrow) of the compact radio 
    counterpart of XTE J1118+480 relative to the extragalactic
    background radio sources Cal. 1 and Cal. 2, measured at
    $\lambda$2cm (blue symbols) and $\lambda$3.6cm (red symbols) on
    2000 May 4 (triangles) and 2000 July 24 (squares). The black
    crosses mark the mean positions.  For clarity, the J2000
    equatorial coordinates (black lines) and galactic coordinates
    (dashed brown lines) are shown in expanded scales of 2$^{\circ}$
    and 1$^{\circ}$ per division respectivelly.  The position of Ref
    1, the primary reference source is identical at both wavelengths
    and both epochs, by definition.  The position of Ref. 2 is the
    same at all wavelengths and epochs, within rms $\sim$0.35 mas. To
    relate the astrometric positions of XTE J1118+480 in Table 1 to
    the galactic frame, a correction for the change in parallax was
    applied.  The position of XTE J1118+480 shifts by -16.8$\pm$1.6
    mas yr$^{-1}$ in right ascension and -7.4$\pm$1.6 mas yr$^{-1}$ in
    declination at both wavelengths.  The thin black arrow shows the
    proper motion of the optical companion measured from a set of four
    photographic plates that cover a time span of $\sim$43 years (from
    1953 to 1996).  A linear fit to the object positions yielded a
    proper motion of $-10.5 \pm 3.2$ mas yr$^{-1}$ in right ascension
    and of $-5.9 \pm 3.2$ mas yr$^{-1}$ in declination, after
    correcting for the galactic rotation and for the peculiar motion
    of the Sun. The difference in position angle between the radio and
    optical is $6^{\circ}$, and the errors $\pm5^{\circ}$ and
    $\pm15^{\circ}$, respectivelly.  }
\end{figure*}

\begin{figure*}[p]
  {\par\centering \resizebox*{1\textwidth}{!}{
      \rotatebox{90}{\includegraphics{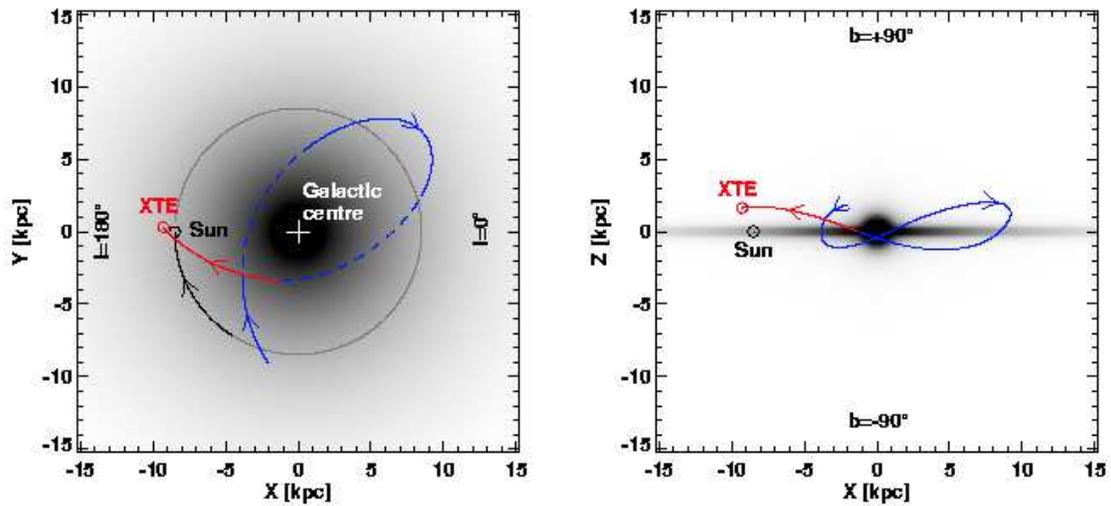}}}
    \par}
  \caption{Galactic orbit of XTE J1118+480 during the last 
    orbital period of the Sun around the Galactic centre (240 Myr).
    The mid-plane mass density distribution of the Galactic bulge and
    disk are represented by a linear grey scale. The last section of
    the orbit since the source left the plane 37.3$\pm$4.8 Myr ago at
    a galactocentric distance of 3.8$\pm$0.5 kpc is in red color. The
    trajectory of the Sun during the later time is indicated by the
    thick black arc. The source left the plane towards the Northern
    Galactic Hemisphere with a galactocentric velocity of 348$\pm$18
    km s$^{-1}$, which after substraction of the velocity vector due
    to galactic rotation, corresponds to a peculiar space velocity of
    217$\pm$18 km s$^{-1}$ relative to the galactic disk frame, and a
    component perpendicular to the plane of 126$\pm$18 km s$^{-1}$.
    The orbit of XTE J1118+480 has an eccentricity e=0.54 and is
    similar to the orbits of halo objects, such as globular clusters.
    In the oscillating motion about the plane it has just turn around
    to fall back in. At the present epoch XTE J1118+480 is 
    at a distance from the Sun of only 1.85$\pm$0.36 kpc flying through 
    the Galactic local neighborhood with a velocity of 145 km s$^{-1}$. 
    {\bf Left:} View from above the Galactic plane; {\bf Right:} side view}
    \end{figure*}

\end{document}